\documentclass[twocolumn,psfig,aps,showpacs]{revtex4}
\usepackage{amsfonts}
\usepackage{amsmath}
\usepackage{graphicx}
\usepackage{bm}
\usepackage{amssymb}
\usepackage{dcolumn}
\usepackage{psfig}

\begin{document}

\title{Thermodynamics of Spin-$1/2$ AF-AF-F and F-F-AF Trimerized Quantum
Heisenberg Chains}
\author{Bo Gu}
\author{Gang Su$^{\ast }$}
\affiliation{College of Physical Sciences, Graduate University of Chinese Academy of
Sciences, P. O. Box 4588, Beijing 100049, China}
\author{Song Gao}
\affiliation{College of Chemistry and Molecular Engineering, State Key Laboratory of Rare
Earth Materials Chemistry and Applications, Peking University, Beijing
100871, China}

\begin{abstract}
The magnetization process, the susceptibility and the specific heat of the
spin-$1/2$ AF-AF-F and F-F-AF trimerized quantum Heisenberg chains have been
investigated by means of the transfer matrix renormalization group (TMRG)
technique as well as the modified spin-wave (MSW) theory. A magnetization
plateau at $m=1/6$ for both trimerized chains is observed at low
temperature. The susceptibility and the specific heat show various behaviors
for different ferromagnetic and antiferromagnetic interactions and in
different magnetic fields. The TMRG results of susceptibility and the
specific heat can be nicely fitted by a linear superposition of double
two-level systems, where two fitting equations are proposed. Three branch
excitations, one gapless excitation and two gapful excitations, for both
systems are found within the MSW theory. It is observed that the MSW theory
captures the main characteristics of the thermodynamic behaviors at low
temperatures. The TMRG results are also compared with the possible
experimental data.
\end{abstract}

\maketitle

\section{INTRODUCTION}

Physical properties of low-dimensional magnetic materials have attracted
considerable attention at the border of condensed matter physics. In
particular, the bond-alternating quantum magnets with both ferromagnetic (F)
and antiferromagnetic (AF) couplings have been extensively investigated,
because they have exhibited quite fascinating properties owing to the
competition of F and AF interactions. A simple AF bond-alternating chain
(BAC) may be a F-AF alternating quantum Heisenberg chain, and the spin-1/2
copper-based compound $(IPA)CuCl_{3}$\cite{F-AF} is a typical example, in
which the Haldane-like behavior has been observed. This kind of alternating
chains have been actively studied both theoretically and experimentally in
the last decades. Another interesting antiferromagnetic BAC is the F-F-AF
alternating Heisenberg chain with period $n=3$. According to
Oshikawa-Yamanaka-Affleck (OYA)\cite{OYA}, the F-F-AF chain with spin $S$
may exhibit a magnetization plateau at $m=S/3$, where $m$ is the
magnetization per site. In a typical example of spin-1/2 F-F-AF chain of $%
3CuCl_{2}\cdot 2dx$ with strong F coupling ($J_{F}$) and weak AF coupling ($%
J_{AF} $), however, no plateau in the low-temperature magnetic curve has
been observed\cite{F-F-AF}. Such a breakdown of the magnetization plateau in
small bond ratio of $|J_{AF}/J_{F}|$\cite{Breakdown} suggests that the OYA
condition is only a necessary condition.

There is another intriguing BAC with period $n=3$, which consists of AF-AF-F
alternations, and is a ferrimagnet. The examples of AF-AF-F chains are $%
[Mn(L)_{2}(N_{3})_{2}]_{n}$ with spin $S=5/2$\cite{AF-AF-F-1}, $%
[M(4,4^{\prime }bipy)(N_{3})_{2}]_{n}$ with $M=Co$ $(S=3/2)$ and $Ni$ $(S=1)$%
\cite{AF-AF-F-2}, {\it and $[Mn(N_{3})_{2}(bpee)]_{n}$ with $S=5/2$\cite%
{AF-AF-F-3}}. These compounds have strong F and weak AF couplings, and no
magnetization plateaus have been seen experimentally. If the ratio $%
|J_{AF}/J_{F}|$ are large enough, the plateaus could be expected during the
magnetizing process. We note that the plateau has been observed
experimentally in a F-F-AF-AF BAC compound, $Cu(3-Clpy)_{2}(N_{3})_{2}$ \cite%
{F-F-AF-AF,Hagiwara}, which possesses a relatively large ratio $%
|J_{AF}/J_{F}|$ \cite{Minami, Xiang}.

Inspired by the exotic magnetic properties of the BAC compounds with period $%
n=3$ observed experimentally, in this paper, we shall invoke the transfer
matrix renormalization group (TMRG) technique \cite{TMRG} to study the
thermodynamic properties of the AF-AF-F chain and F-F-AF chain,
respectively. The rest of this paper is organized as follows. In Sec. II, we
shall construct the model Hamiltonian for the trimerized J-J-J$^{\prime }$
Heisenberg spin chain. In Sec. III, we shall present our TMRG results on the
thermodynamic behaviors of the systems. The comparisons between the TMRG
results and the results of the modified spin-wave (MSW) theory will be
discussed in Sec. IV. Finally, a brief summary will be given.

\section{MODEL}

\begin{figure}[tbp]
\vspace{0.5cm} \includegraphics[width = 8.5cm]{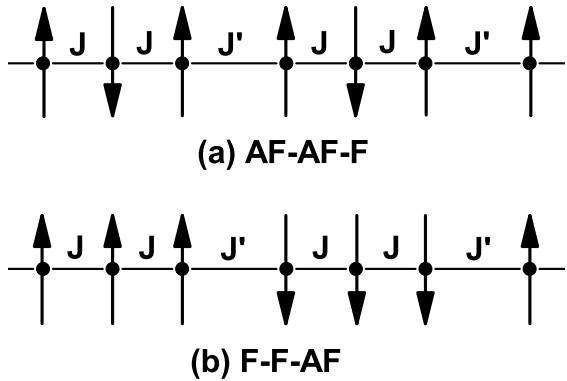}
\caption{ Spin arrangements of the J-J-J$^{\prime }$ trimerized spin chain
with $S=1/2$, where $\uparrow $ denotes spin up, $\downarrow $ denotes spin
down, $J$ and $J^{\prime }$ are exchange couplings. (a) AF-AF-F chain, $%
J=J_{AF}$, $J^{\prime }=J_{F}$; (b) F-F-AF chain, $J=J_{F}$, $J^{\prime
}=J_{AF}$. }
\label{chain}
\end{figure}

Let us consider a trimerized $S=1/2$ J-J-J$^{\prime }$ Heisenberg quantum
spin chain. The Hamiltonian of the system reads 
\begin{eqnarray}
H &=&\sum\limits_{j}(J\mathbf{S}_{3j-2}\cdot \mathbf{S}_{3j-1}+J\mathbf{S}%
_{3j-1}\cdot \mathbf{S}_{3j}+J^{\prime }\mathbf{S}_{3j}\cdot \mathbf{S}%
_{3j+1})  \notag \\
&&-h\sum\limits_{j}S_{j}^{z},  \label{Hamilton}
\end{eqnarray}%
where $J$ and $J^{\prime }$ are exchange integrals with $J$, $J^{\prime }>0$
denoting the antiferromagnetic coupling and $J$, $J^{\prime }<0$ the
ferromagnetic coupling, $h$ is the external magnetic field, and we take $%
g\mu _{B}=1$. The schematic spin arrangements of the system are shown in
Figs. \ref{chain}. When $J=J^{\prime }=J_{F}$ (or $J_{AF}$), the system
becomes a uniform $S=1/2$ Heisenberg ferromagnetic (or antiferromagnetic)
chain. 

When $J\neq J^{\prime }$, as indicated in Figs. \ref{chain}, if $J=J_{AF}$, $%
J^{\prime }=J_{F}$, it is a ferrimagnetic spin chain (i.e. the configuration
is AF-AF-F); if $J=J_{F}$, $J^{\prime }=J_{AF}$, it becomes an
antiferromagnetic system (i.e. the configuration is F-F-AF). The competition
of F and AF couplings in these two trimerized chains would give rise to
plenty of interesting characteristics. The magnetic properties of this model
with S=1/2, 1, 3/2 and 2 in a magnetic field at zero temperature have been
investigated by use of the density matrix renormalization group method\cite%
{Gubo}. It is purpose of this paper to study the thermodynamic properties of
these trimerized quantum Heisenberg spin chains by means of the TMRG
technique\cite{TMRG}.

\section{TMRG RESULTS}

The TMRG method was detailed in two nice reviews\cite{Review}, and we shall
not repeat the technical skills here for concise. In our calculations, the
number of kept optimal states is taken as $m=64$ for the susceptibility $%
\chi $, and $m=80$ for the specific heat $C$, where the width of the
imaginary time slice is taken as $\varepsilon =0.1$. We have used different $%
m$ and $\varepsilon $ to verify the accuracy of calculations. At high
temperature, the error caused by the Trotter-Suzuki decomposition is
important and is of the order $\varepsilon ^{3}$ for a fixed $m$. At low
temperature, the error resulting from the basis truncation becomes
important, which drops exponentially with increasing $m$ initially and
reaches to a finite value. The Trotter-Suzuki error is less than $10^{-3}$,
and the truncation error is smaller than $10^{-6}$ in our calculations. The
physical quantities presented below are calculated down to $T=0.025$ (in
units of $J_{F}$).

\subsection{MAGNETIZATION}

\begin{figure}[tbp]
\centering\includegraphics[width = 8.5cm]{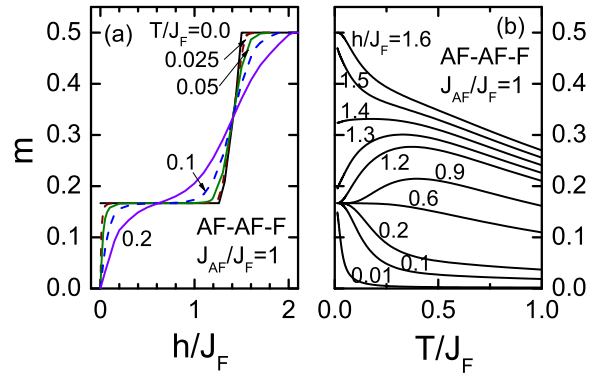}
\caption{(Color online) For the AF-AF-F chain with $J_{AF}/J_{F}=1$: (a) The
magnetization per site $m$ as a function of magnetic field $h$ at different
temperatures; (b) The temperature dependence of $m$ under different magnetic
fields.}
\label{mh-aaf}
\end{figure}

Fig. \ref{mh-aaf}(a) shows the magnetization per site $m$ as a function of
magnetic field $h$ at different temperatures for the AF-AF-F chain with $%
J_{AF}/J_{F}=1$. At zero temperature, a magnetization plateau at $m=1/6$ was
observed\cite{Gubo}. As illustrated in Fig. \ref{chain}(a), the ground state
of the $S=1/2$ AF-AF-F chain in the absence of the magnetic field
corresponds to $m=1/6$. When the external field is applied, the ground state
with $m=1/6$ still persists for $h/J_{F}<1.3$, which is consistent with the
OYA condition\cite{OYA}. For $h/J_{F}>1.3$ the plateau state is destructed,
and $m(h)$ increases rapidly with increasing the field. When the field is
increased to $h/J_{F}=1.6$, $m(h)$ becomes saturated, and the ground state
becomes a fully polarized state. When the temperature is increased, the
magnetization plateau retains at low temperature, and is gradually smeared
out at high temperature owing to the thermal fluctuations. It is noting that
the magnetization $m(h)$ as a function of $h$ at finite temperature starts
from zero, that is nothing but the result of Mermin-Wagner theorem.

Fig. \ref{mh-aaf}(b) gives the temperature dependence of the magnetization $%
m(T)$ under different magnetic fields for the AF-AF-F chain with $%
J_{AF}/J_{F}=1$. It can be seen that $m(T)$ behaves distinctly for different
ranges of the external field. At $h<0.6$, $m(T)$ first decreases rapidly,
and then declines slowly with increasing temperature; at $0.6<h<1.4$, $m(T)$
first increases, and then declines slowly with increasing temperature; at $%
h>1.4$, $m(T)$ declines with increasing temperature. At $h=0.6$ and $1.4$, $%
m(T)$ appears to remain constant with zero curvature at low temperature, and
declines slowly with temperature. It is found that just below and above $%
h=0.6$ and $1.4$, the curvature of $m(T)$ changes abruptly with opposite
signs, implying that $h=0.6$ and $1.4$ could be viewed as the crossover
fields. This observation is also consistent with Fig. \ref{mh-aaf}(a) where
there are two crossing points in the curves for different temperatures at $%
h=0.6$ and $1.4$.

\begin{figure}[tbp]
\vspace{0.5cm} \includegraphics[width = 8.5cm]{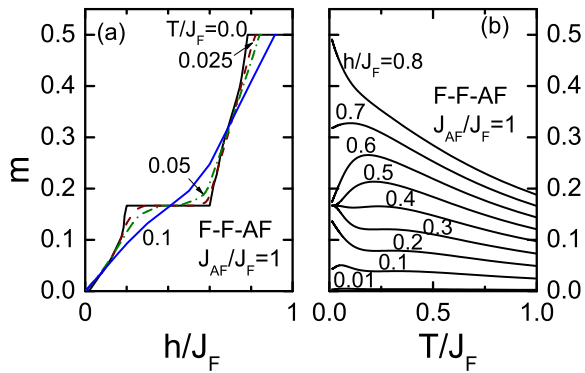}
\caption{(Color online) For the F-F-AF chain with $J_{AF}/J_{F}=1$: (a) The
magnetization per site $m$ as a function of the magnetic field at different
temperatures; (b) The temperature dependence of $m$ under different fields.}
\label{ffa-m}
\end{figure}

Fig. \ref{ffa-m}(a) shows the field dependence of the magnetization per site 
$m(h)$ at different temperatures for the F-F-AF chain with $J_{AF}/J_{F}=1$.
A magnetization plateau at $m=1/6$ is also observed at zero temperature. As
indicated in Fig. \ref{chain}(b), the ground state of the $S=1/2$ AF-AF-F
chain in the absence of a magnetic field corresponds to $m=0$. With
increasing the magnetic field, $m$ first increases till to $m=1/6$ at $%
h/J_{F}\approx 0.2$, then goes into the plateau state with $m=1/6$ for $%
h/J_{F}<0.7$, beyond which the plateau state is destroyed, and $m(h)$ then
increases rapidly to a saturation at $h/J_{F}=0.8$ where the system becomes
fully polarized. The occurrence of the plateau state with $m=1/6$ is also
consistent with the OYA condition\cite{OYA}. It is seen that at low
temperature the magnetization plateau still persists, though the width of
plateau is getting narrow. At high temperature, the plateau is gradually
smeared out by thermal fluctuations, and m(h) increases nonmonotonically
till the saturation with increasing the magnetic field.

The temperature dependence of the magnetization per site $m(T)$ under
different fields for the F-F-AF chain with $J_{AF}/J_{F}=1$ is presented in
Fig. \ref{ffa-m}(b). One may see that $m(T)$ shows different behaviors at
low temperature in different ranges of the magnetic field. At $h<0.2$, $m(T)$
increases first and then declines smoothly with increasing temperature; at $%
0.2<h<0.4$, $m(T)$ decreases dramatically at low temperature and then slowly
with increasing temperature; at $0.4<h<0.7$, $m(T)$ increases at low
temperature and then declines slowly with increasing temperature; at $h>0.7$%
, $m(T)$ declines remarkably with increasing temperature. At $h=0.4$ and $%
0.7 $, $m(T)$ remains almost unchanged at lower temperature, and then
declines slowly with increasing $T$. Again, the curvature of $m(T)$ changes
abruptly with opposite signs just below and above $h=0.4$ and $0.7$. The
results presented in Fig. \ref{ffa-m}(b) suggest that at low temperature the
system may enter into different states in different ranges of the magnetic
field, namely, for $h<0.2$, the system is in an ordering state; for $%
0.2<h<0.4$ the system could go into the spin-canting state; for $0.4<h<0.7$
the system enters into the plateau state where the excitations are gapful;
and for $h>0.7$ the system goes into another ordering state before
saturation.

It is interesting to note that the magnetization as a function of magnetic
field and temperature exhibits different behaviors for the AF-AF-F chain and
the F-F-AF chain, for the former is a ferrimagnet, while the latter is an
antiferromagnet. However, both systems shows a magnetization plateau with $%
m=1/6$ at low temperature, being in agreement with the OYA condition.

\subsection{SUSCEPTIBILITY}

\begin{figure}[tbp]
\centering\includegraphics[width = 8.5cm]{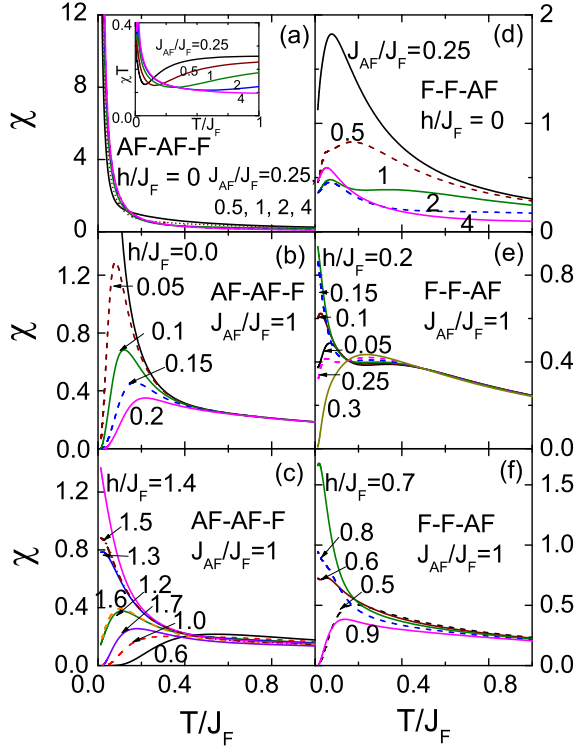}
\caption{(Color online) The temperature dependence of the susceptibility $%
\protect\chi $. For the AF-AF-F chain (a) at $h=0$ for different $%
J_{AF}/J_{F}$, (b) and (c) at $J_{AF}/J_{F}=1$ for different external
fields, where the inset of (a) depicts $\protect\chi T$ against $T$ for
different $J_{AF}/J_{F}$. For the F-F-AF chain (d) at $h=0$ for different $%
J_{AF}/J_{F}$, (e) and (f) at $J_{AF}/J_{F}=1$ for different external
fields. }
\label{xt}
\end{figure}

Let us now present the TMRG results of the susceptibility ($\chi $) as a
function of temperature ($T$) for the AF-AF-F chain. In the absence of the
magnetic field, the susceptibility diverges when the temperature tends to
zero, and with increasing temperature, $\chi $ decreases rapidly, as shown
in Fig. \ref{xt}(a). For different $J_{AF}/J_{F}$, $\chi (T)$ exhibits
similar behaviors at low $T$, but distinct behaviors at moderate $T$, as
manifested in the inset of Fig. \ref{xt}(a), where $\chi T$ versus $T$ is
plotted for various $J_{AF}/J_{F}$, and a low-$T$ divergence and a round
minimum are observed for $J_{AF}/J_{F}\leq 1$, indicating that when the F
interaction $J_{F}$ predominates, the system shows a ferrimagnetic behavior;
when the AF interaction $J_{AF}$ is predominant, the system falls into a
trimerized bond state. In the presence of the magnetic field, the
susceptibility $\chi $ shows quite different behaviors, as presented in
Figs. \ref{xt}(b) and (c) at $J_{AF}/J_{F}=1$. Unlike the case of $h=0$, $%
\chi $ under nonzero $h$ exhibits a peak at low temperature except $h=0.6$
and $1.4$ where with increasing $T$, $\chi $ first increases and then drops
slowly for the former, and goes to divergence at low temperature for the
latter. With the increase of $h$, the peak of $\chi $ appears to become
rounded, and the peak height of $\chi $ gets suppressed for $h<0.6$, then
increased for $0.6<h<1.4$, and then becomes suppressed again for $h>1.4$.
These observations show again that the system can enter into different
states within different ranges of the magnetic field.

For the F-F-AF chain, the temperature dependence of the susceptibility $\chi 
$ for different $J_{AF}/J_{F}$ and $h$ is shown in Figs. \ref{xt}(d)-(f),
respectively. In the absence of the magnetic field, the susceptibility
reveals peaks at low temperature for different $J_{AF}/J_{F}$, and $\chi =0$
at $T=0$. When $J_{AF}/J_{F}$ is either smaller or larger, e.g., $%
J_{AF}/J_{F}=0.25$, $2$, $4$, $\chi $ shows one peak; while $%
J_{AF}/J_{F}=0.5 $ and $1$, $\chi $ shows a double-peak structure, as
manifested in Fig. \ref{xt}(d). The calculated results indicate that the
system shows an AF behavior. It seems that the double-peak structure comes
from the two different kinds of excitations induced by the competition
between F and AF interactions. In the presence of the magnetic field, the
temperature dependence of the susceptibility $\chi $ at $J_{AF}/J_{F}=1$ for
different external fields are presented in Figs. \ref{xt}(e) and (f). It is
found that $\chi (T)$ behaves variously in different ranges of the magnetic
field. When $h$ is small, $\chi (T)$ shows a double-peak structure at low
temperature with the first peak sharp and the second peak round; with
increasing $h$, the height of the first peak grows dramatically, while that
of the second round peak leaves almost unchanged; when $h/J_{F}>0.2$, the
first peak is remarkably suppressed at $h/J_{F}=0.25$, and vanishes at $%
h/J_{F}=0.3$, while the second round peak becomes slightly higher, as shown
in Fig. \ref{xt}(e). When $h/J_{F}>0.5$, the second peak of $\chi $ moves to
the low-temperature side, and at $h/J_{F}=0.7$, the two peaks merge into a
single; when $h$ is increased further, the peak of $\chi $ is suppressed
again. The behaviors of $\chi (T)$ show that the F-F-AF chain can enter into
different states at low temperatures under different magnetic fields,
consistent with the results presented in Fig. \ref{ffa-m}(b).

\begin{figure}[tbp]
\vspace{0.5cm} \includegraphics[width = 8.5cm]{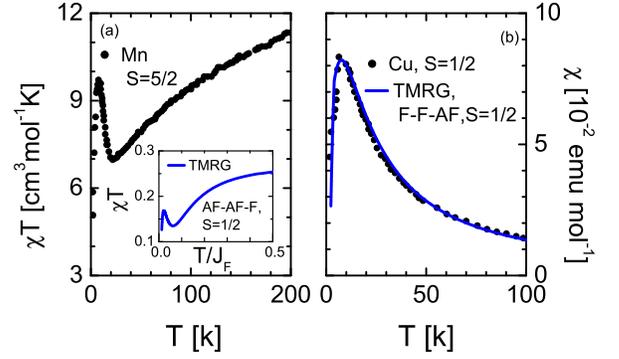}
\caption{(Color online) (a) The experimental data of $\protect\chi T$ of the
spin-5/2 AF-AF-F chain $[Mn_{3}(N_{3})_{6}(bpe)_{3}]$ (from Ref.\protect\cite%
{AF-AF-F-3}); Inset presents the TMRG result of an AF-AF-F chain with $S=1/2$%
. (b) A comparison of the susceptibility for $3CuCl_{2}\cdot 2dx$ with the
TMRG result of a F-F-AF chain with $S=1/2$, where the experimental data are
taken from Ref. \protect\cite{F-F-AF}.}
\label{theory-tmrg}
\end{figure}

To verify our TMRG results, we have included the experimental data of an
AF-AF-F chain compound $[Mn_{3}(N_{3})_{6}(bpe)_{3}]$ and a F-F-AF chain
compound $3CuCl_{2}\cdot 2dx$ in Figs. \ref{theory-tmrg} for a comparison.
Fig. \ref{theory-tmrg}(a) shows the experimental data of $\chi T$ versus $T$
for $[Mn_{3}(N_{3})_{6}(bpe)_{3}]$, whose magnetic structure is an AF-AF-F
chain with spin-$5/2$ \cite{AF-AF-F-3}. Owing to our computing capacity, we
cannot at present calculate directly the magnetic properties of the spin-$%
5/2 $ AF-AF-F chain by using the TMRG method. As a qualitative
understanding, however, in the inset of Fig. \ref{theory-tmrg}(a), we have
presented our TMRG result of a spin-$1/2$ AF-AF-F chain with $h/J_{F}=0.003$
and $J_{AF}/J_{F}=0.2$. It can be seen that the experimental and calculated
curves share qualitatively similar characteristics. Fig. \ref{theory-tmrg}%
(b) gives a comparison of the experimental susceptibility of the compound $%
3CuCl_{2}\cdot 2dx$, which is a F-F-AF chain with spin-$1/2$ \cite{F-F-AF},
with the TMRG result. It is found that our TMRG result fits well with the
experimental data, with the coupling parameters $J_{AF}/J_{F}=0.2$, $%
J_{F}=100$ $K$, and $h/J_{F}=0.005$. Our calculated result is in agreement
with the previous theoretical result\cite{F-F-AF}.

\subsection{SPECIFIC HEAT}

\begin{figure}[tbp]
\centering\includegraphics[width = 8.5cm]{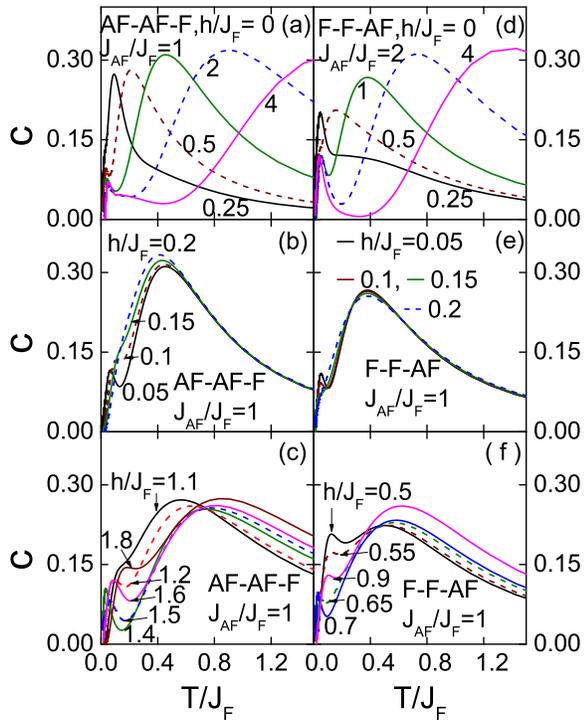}
\caption{(Color online) The temperature dependence of the specific heat $C$.
For the AF-AF-F chain (a) at $h=0$ for different $J_{AF}/J_{F}$, (b) and (c)
at $J_{AF}/J_{F}=1$ for different external fields. For the F-F-AF chain (d)
at $h=0$ for different $J_{AF}/J_{F}$, (e) and (f) at $J_{AF}/J_{F}=1$ for
different external fields. }
\label{ct}
\end{figure}

In this subsection, we shall discuss the specific heat of the spin-1/2
AF-AF-F and F-F-AF chains, as shown in Figs. \ref{ct} for different $%
J_{AF}/J_{F}$ and $h$. For the AF-AF-F chain, in absence of the magnetic
field, the specific heat $C$ as a function of temperature shows clearly a
double-peak structure, as indicated in Fig. \ref{ct}(a). With increasing $%
J_{AF}/J_{F}$, the low-temperature peak of $C$ leaves almost unchanged,
while the second peak moves to the high-temperature side, and gets rounded.
This feature of the specific heat shows that the system might exist two
different kinds of excitations due to the competition of F and AF
interactions, as will be discussed in next section. It seems that the first
peak corresponds to the F excitations, while the second peak corresponds to
the AF excitations.

In presence of the magnetic field, the temperature dependence of the
specific heat C for the AF-AF-F chain with $J_{AF}/J_{F}=1$ is shown in
Figs. \ref{ct}(b) and (c) under different fields. When $h$ is small, the
double-peak structure of $C$ is observed; with increasing $h$, the first
peak tends to vanish, while the second peak remains almost unchanged, as
shown in Fig. \ref{ct}(b). When $0.2<h/J_{F}<1.1$, the two peaks become a
single. When $h/J_{F}>1.1$, as presented in Fig. \ref{ct}(c), the first peak
recovers anew, and with the increase of $h$, the first peak moves to the
lower temperature side till $h/J_{F}=1.4$, then shifts to the high
temperature side for $1.8>h/J_{F}>1.4$, and tends to vanish for $h/J_{F}>1.8$%
, while the second peak shifts slightly to the higher temperature side with
increasing $h$. The results show that the system is in a few different
states under different magnetic fields.

For the F-F-AF chain, the temperature dependence of the specific heat $C$ is
shown in Figs. \ref{ct}(d)-(f) for different $J_{AF}/J_{F}$ and $h$. In
absence of the magnetic field, $C$ also shows a double-peak structure; with
increasing $J_{AF}/J_{F}$, the height of the first peak of $C$ is suppressed
but its position leaves unchanged, while the second peak tends to shift to
the high-temperature side, as illustrated in Fig. \ref{ct}(d). It is seen
that the qualitative behavior is similar to that of the AF-AF-F chain
presented in Fig. \ref{ct}(a). The results suggest that the F-F-AF chain has
also two kinds of excitations due to the competition of F and AF
interactions.

Figs. \ref{ct}(e) and (f) gives the temperature dependence of the specific
heat $C$ for the F-F-AF chain with $J_{AF}/J_{F}=1$ under different magnetic
fields. Similar to the AF-AF-F chain, with increasing $h$, the first peak of
C is gradually suppressed, while the second peak retains nearly intact till $%
h/J_{F}=0.2$. When $0.2<h/J_{F}<0.5$, the first peak is completely
suppressed; when $0.5<h/J_{F}<0.7$, with the increase of $h$, the first peak
recovers again with the peak height decreased and the peak position shifted
to the lower temperature side, while the second peak moves slightly to the
high temperature side with the peak height a little bit enhanced; when $%
h/J_{F}>0.7$, the situation changes, namely, with increasing $h$, the first
peak of $C$ is enhanced and moves to the high temperature side, and the
second peak is also promoted. The present observation displays that the
F-F-AF chain falls into different thermodynamic states under different
external fields, and has some behaviors similar to the AF-AF-F chain,
although both chains have quite different ground states, as the latter is a
ferrimagnet, while the former is an antiferromagnet.

\subsection{FITTING\ TO\ A SUPERPOSITION OF DOUBLE TWO-LEVEL SYSTEMS}

The double-peak structure of the temperature dependence of the specific heat
can be viewed as a minimum at low temperature and a Schottky-like maximum at
high temperature. The Schottky-like anomaly can be fitted with a so-called
two-level system, where the thermal population of the levels is governed by
the Maxwell-Bolzman statistics, $n_{i}=e^{-\varepsilon _{i}/k_{B}T}/Z$ ($%
i=1,2$) with the distribution function $Z=e^{-\varepsilon
_{1}/k_{B}T}+e^{-\varepsilon _{2}/k_{B}T}$, and the total energy $%
E(T)=n_{1}\varepsilon _{1}+n_{2}\varepsilon _{2}$. The specific heat can be
obtained by $C=(\partial E/\partial T)$. Define $\varepsilon _{1}=0$, $%
\varepsilon _{2}=\Delta $. The Schottky-like specific heat can be written as 
\begin{equation}
C_{s}(\Delta ,T)=k_{B}(\frac{\Delta }{k_{B}T})^{2}\frac{e^{\Delta /k_{B}T}}{%
(1+e^{\Delta /k_{B}T)^{2}}}.  \label{C-twolevel}
\end{equation}%
Similarly, it is found that the magnetic susceptibility can be fitted by\cite%
{Mohn}, 
\begin{equation}
\chi _{s}(\Delta ,T)=k_{B}(\frac{\Delta }{k_{B}T})\frac{e^{\Delta /k_{B}T}}{%
(1+e^{\Delta /k_{B}T)^{2}}},  \label{x-twolevel}
\end{equation}%
where $\Delta $ is the excitation gap. However, our TMRG results cannot be
well fitted by these above two formulae. By noting that the present systems
such as the AF-AF-F and F-F-AF chains are quantum trimerized systems with a
period $n=3$, and the double-peak structure may be regarded as the
consequence of the competition between F and AF excitations, the
thermodynamic properties of these systems could be mimicked by a
superposition of double two-level systems. As a result, the susceptibility
for the AF-AF-F and F-F-AF chains may be expected to fit by 
\begin{equation}
\chi (T)=A_{1}\chi _{s}(\Delta _{1},T)+A_{2}\chi _{s}(\Delta _{2},T),
\label{x-double}
\end{equation}%
and the double-peak structure of the specific heat can be fitted by the
following form: 
\begin{equation}
C(T)=A_{1}C_{s}(\Delta _{1},T)+A_{2}C_{s}(\Delta _{2},T),  \label{C-double}
\end{equation}%
where $A_{1}$, $A_{2}$ are fitting parameters, and $\Delta _{1}$, $\Delta
_{2}$ are the excitation gaps for the double two-level systems. In these
above equations, $\chi _{s}$ and $C_{s}$ are defined by Eqs. (\ref%
{x-twolevel}) and (\ref{C-twolevel}), respectively.

\begin{figure}[tbp]
\vspace{0.5cm} \includegraphics[width = 8.5cm]{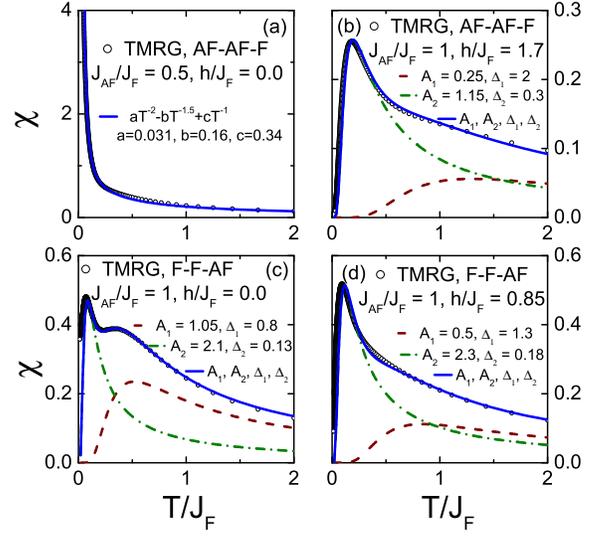}
\caption{(Color online) The TMRG results of the susceptibility as a function
of temperature are fitted. For the AF-AF-F\ chain (a) by $%
a(T/J_{F})^{-2}-b(T/J_{F})^{-3/2}+c(T/J_{F})^{-1}$ for $h=0$, (b) by Eq. (%
\protect\ref{x-double}) for $h$ nonzero; for the F-F-AF chain by Eq. (%
\protect\ref{x-double}) for (c) $h=0$, and (d) $h$ nonzero.}
\label{3l-xt}
\end{figure}

For the AF-AF-F chain, in absence of the magnetic field, we have found that $%
\chi (T)$ cannot be fitted by Eq. (\ref{x-double}), but can be well fitted
by the equation $\chi (T)=a(T/J_{F})^{-2}-b(T/J_{F})^{-3/2}+c(T/J_{F})^{-1}$
with $a$, $b$, and $c$ the fitting parameters, as shown in Fig. \ref{3l-xt}%
(a) for $J_{AF}/J_{F}=0.5$ as an example. We note that the susceptibility of
a spin-$1/2$\ ferromagnetic system was expanded as $\chi J/L(g\mu
_{B})^{2}=0.04167t^{-2}-0.145t^{-1.5}+0.17t^{-1}+O(t^{-0.5})$ with $%
t=-k_{B}T/J$ $(J<0)$ at low temperatures\cite{Taka1}. Thus, it is not
surprising that the present system shows a similar behavior, as the AF-AF-F
chain is a ferrimagnet. However, in presence of the magnetic field, the
susceptibility $\chi (T)$ of the AF-AF-F chain can be nicely fitted by Eq. (%
\ref{x-double}), as compared in Fig. \ref{3l-xt}(b) for $J_{AF}/J_{F}=1$ and 
$h/J_{F}=1.7$ as an example. Figs. \ref{3l-xt}(c) and (d) give the TMRG
fitting results in terms of Eq. (\ref{x-double}) for the susceptibility $%
\chi (T)$ of the F-F-AF chain in absence and presence of the magnetic field,
respectively. One may see that apart from the case where $\chi (T)$ is
divergent at lower temperature and cannot be fitted by Eq. (\ref{x-double}),
the susceptibility as a function of temperature for both trimerized chains
can be well fitted by a superposition of the double two-level systems.

\begin{figure}[tbp]
\vspace{0.5cm} \includegraphics[width = 8.5cm]{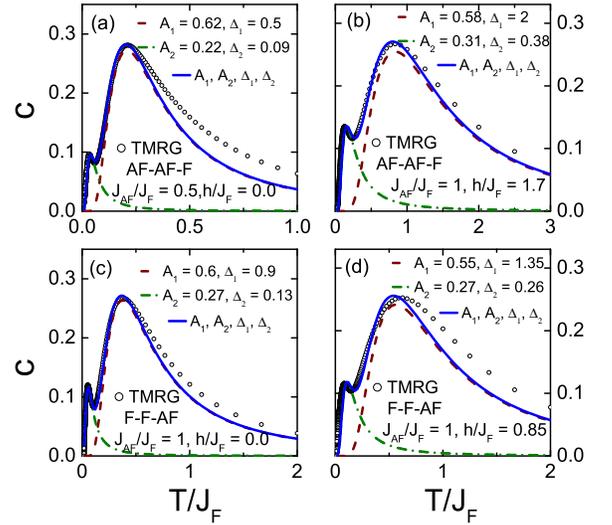}
\caption{(Color online) The TMRG results of the specific heat as a function
of temperature are fitted by Eq. (\protect\ref{C-double}). For the AF-AF-F
chain with (a) $J_{AF}/J_{F}=0.5$ and $h=0$, and (b) $J_{AF}/J_{F}=1.7$ and $%
h/J_{F}=1.7$; for the F-F-AF chain with $J_{AF}/J_{F}=1$ and (c) $h=0$ and
(d) $h/J_{F}=0.85$.}
\label{3l-ct}
\end{figure}

Figs. \ref{3l-ct} present the TMRG results of the specific heat as a
function of temperature fitting to a superposition of double two-level
systems characterized by Eq. (\ref{C-double}) for the AF-AF-F and F-F-AF
chains. It can be found that the specific heat with double-peak structure
can be well fitted by Eq. (\ref{C-double}) at low temperature, with only a
slightly quantitative deviation at high temperature, showing that the main
features of the specific heat of both trimerized chains can be reproduced by
a superposition of double two-level systems. The fitting results could give
the two excitation gaps. The reasons for such nice fittings come from the
fact that these two trimerized chains have three branches of excitations
(see next section), one gapless F excitations, and two gapful excitations,
which could be equivalently treated by a superposition of double two-level
systems with two gaps $\Delta _{1}$ and $\Delta _{2}$. We have also found
that a direct fitting to a three-level system is not as good as to double
two-level systems.

\section{Modified Spin-Wave Theory}

The zero-field specific heat of a quantum ferrimagnet, which is believed as
an intrinsic double-peak structure of topological origin and is different
from the external field induced double-peak structure, was investigated by
using the modified spin-wave (MSW) theory\cite{Taka2,Nakanishi,Yamamoto},
where it has been found that in the case of the AF-AF-F-F BAC, there are
four distinct branches of spin-wave excitations, say, the lower two
ferromagnetic bands construct the low-temperature bump, while an upper
ferromagnetic band and an antiferromagnetic band contribute to a
Schottky-type peak at mid temperatures.

The field-induced double-peak structure of the specific heat in a quantum
ferrimagnet has also been studied by using the linear spin-wave theory\cite%
{Maisinger}. In the case of spin-$(1,\frac{1}{2})$ ferrimagnet, a simple
picture was given by noting that the Zeeman term introduces a gap to the F
excitations that increases with increasing the field, and a gap to the AF
excitations that decreases with increasing the field, and the specific heat
really reflects the dual structure of the antiferromagnetic and
ferromagnetic excitations for all fields.

The MSW theory can also be applied to the present $S=1/2$ AF-AF-F and F-F-AF
chains in absence of the external field so that our TMRG results of the
thermodynamic quantities may be used to compare with the corresponding MSW
results. In the conventional spin-wave scheme, the spin deviations in each
sublattice diverge in the one-dimensional (1D) antiferromagnets, but the
quantum as well as thermal divergence of the number of bosons can be
overcome in the Takahashi scheme\cite{Taka2} that will be applied to the
present antiferromagnetic F-F-AF chain. The AF-AF-F chain is a ferrimagnet,
whose magnetization should be nonzero in the ground state but zero at finite
temperature, leading to that we can apply the Yamamoto scheme\cite%
{Yama4,Yama5}, where the Lagrange multiplier was introduced directly in the
free energy, to our present ferrimagnetic AF-AF-F chain. The detail
derivations of the MSW formalism are collected in Appendix A, where the
linear modified spin-wave (LMSW) theory, which is up to the order of $%
O(S^{1})$, and the perturbational interacting modified spin-wave (PIMSW)
theory, which is up to the order of $O(S^{0})$, are included. Our results
show that the MSW results may describe well the low-temperature behavior of
the system, but they are not good in agreement with the TMRG results at high
temperature.

\begin{figure}[tbp]
\vspace{0.5cm} \includegraphics[width = 8.5cm]{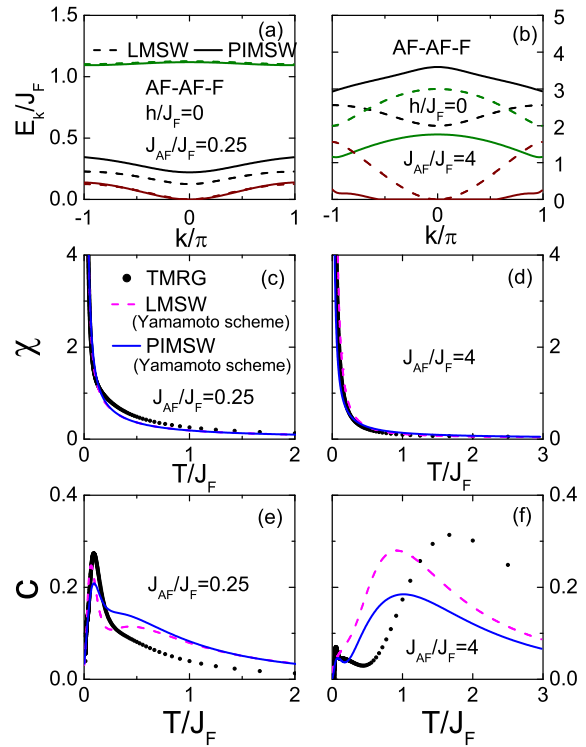}
\caption{(Color online) The excitation spectra of the AF-AF-F chain
calculated by the LMSW and PIMSW methods with (a) $J_{AF}/J_{F}=0.25$ and
(b) $J_{AF}/J_{F}=4$. A comparison of the TMRG results with the LMSW and
PIMSW results of the susceptibility $\protect\chi (T)$ for the AF-AF-F chain
with (c) $J_{AF}/J_{F}=0.25$ and (d) $J_{AF}/J_{F}=4$. The comparison of
TMRG with LMSW and PIMSW results of the specific heat $C(T)$ for the AF-AF-F
chain with (e) $J_{AF}/J_{F}=0.25$ and (f) $J_{AF}/J_{F}=4$.}
\label{msw-aaf}
\end{figure}

Figs. \ref{msw-aaf}(a) and (b) present the zero-field excitation spectra
which were calculated by the LMSW and PIMSW methods within the Yamamoto
scheme for the AF-AF-F chain. It is seen that this trimerized system has
three branches of excitation spectra, one gapless excitation spectrum and
two gapful excitation spectra. When $J_{AF}/J_{F}$ is small, the lower F
gapless spectrum and the upper AF gapful spectrum look coincidence, while
the mid F branches are slightly shifted; when $J_{AF}/J_{F}$ becomes large,
the lower F branch is far separated from the mid and upper excitation
spectra for the PIMSW results, while for the LMSW results, the excitation
spectra are more dispersive, and the mid branch has two crossing points with
the upper branch, showing that the LMSW and PIMSW methods give different
results when the AF interaction is more dominant. Figs. \ref{msw-aaf}(c) and
(d) show the comparison between the TMRG and LMSW and PIMSW results of the
temperature dependence of the zero-field susceptibility for the AF-AF-F
chain with different $J_{AF}/J_{F}$. When $J_{AF}/J_{F}$ is large, the LMSW
and PIMSW results of $\chi (T)$ are in agreement with the TMRG result, as
shown in Figs. \ref{msw-aaf}(d); when $J_{AF}/J_{F}$ is small, say, the F
interaction takes predominant, the low-temperature behavior of $\chi (T)$
calculated by the LMSW and PIMSW theories coincides with the TMRG result,
while the high-temperature behavior shows a somewhat difference, as shown
Figs. \ref{msw-aaf}(c). Figs. \ref{msw-aaf}(e) and (f) give the comparison
of the LMSW and PIMSW results with the TMRG result of the zero-field
specific heat $C(T)$ for the AF-AF-F chain with different $J_{AF}/J_{F}$. It
can be seen that the LMSW and PIMSW theories can be applied to describe the
specific heat of the AF-AF-F chain at lower temperature, where the PIMSW
result can fairly recover the first peak position of $C(T)$ at lower
temperature, but they cannot describe well the high-temperature behavior
where both MSW results look only qualitatively similar to the TMRG result.
However, it can be stated that the double-peak structure of the specific
heat is intimately related to the three branch excitations of the trimerized
system, as manifested by the MSW results.

\begin{figure}[tbp]
\vspace{0.5cm} \includegraphics[width = 8.5cm]{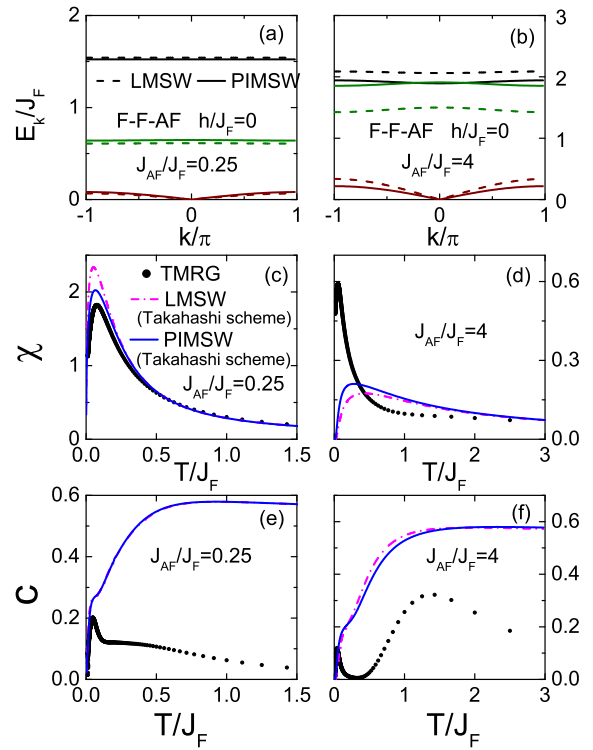}
\caption{(Color online) The excitation spectra of the F-F-AF chain
calculated by the LMSW and PIMSW methods with (a) $J_{AF}/J_{F}=0.25$ and
(b) $J_{AF}/J_{F}=4$. A comparison of the TMRG results with the LMSW and
PIMSW results of the susceptibility $\protect\chi (T)$ for the F-F-AF chain
with (c) $J_{AF}/J_{F}=0.25$ and (d) $J_{AF}/J_{F}=4$. The comparison of
TMRG with LMSW and PIMSW results of the specific heat $C(T)$ for the F-F-AF
chain with (e) $J_{AF}/J_{F}=0.25$ and (f) $J_{AF}/J_{F}=4$. }
\label{msw-ffa}
\end{figure}

Figs. \ref{msw-ffa}(a) and (b) present the excitation spectra calculated by
the LMSW and PIMSW methods within the Takahashi scheme for the F-F-AF chain
with $J_{AF}/J_{F}=0.25$ and $4$ in absence of the magnetic field,
respectively. There are also three branch excitations, i.e., one F gapless
excitation spectrum, and two gapful excitation spectra, which are
responsible for the thermodynamic behaviors of the system at low
temperature. When the AF interaction is weak, the LMSW and PIMSW theories
give a similar result, as shown in Fig. \ref{msw-ffa}(a); when the AF
interaction becomes stronger, both methods produce quite different results
particularly for the upper excitation branches, say, the two upper
excitation spectra calculated from the PIMSW method have two crossings,
while those calculated by the LMSW method are separated, as indicated in
Fig. \ref{msw-ffa}(b), showing that the LMSW and PIMSW methods should be
cautiously applied when the F interaction of the system is weaker. Figs. \ref%
{msw-ffa}(c) and (d) show a comparison between the TMRG and MSW results of
the temperature dependence of the zero-field susceptibility for the F-F-AF
chain with $J_{AF}/J_{F}=0.25$ and $4$, respectively. One may see that when $%
J_{AF}/J_{F}$ is small, the low-temperature peak of $\chi (T)$ can be fairly
recovered, although there are somewhat differences; when $J_{AF}/J_{F}$ is
large, say, the F interaction is weak, the position of low-temperature peak
of $\chi (T)$ can be clearly reproduced, although there are quantitative
changes, indicating that the LMSW and PIMSW methods could capture the main
characteristics of the low-lying excitations of the present trimerized
system at low temperature. Figs. \ref{msw-ffa}(e) and (f) give the
comparison between the TMRG and MSW results of the zero-field specific heat $%
C(T)$ for the F-F-AF chain with $J_{AF}/J_{F}=0.25$ and $4$, respectively.
It is seen that the double-peak structure of $C(T)$ can also be obtained by
the LSMSW and PIMSW methods, which can produce the main features of the
specific heat of the system at low temperature, but have large deviations at
high temperature.

\section{SUMMARY}

In this paper, we have numerically studied the thermodynamics of the $S=1/2$
AF-AF-F and F-F-AF trimerized quantum Heisenberg chains by means of the TMRG
method and the MSW theory. The magnetization process, the low-lying
excitations, the susceptibility and the specific heat of both systems have
been explored. It is found that the TMRG results of the temperature
dependence of the susceptibility as well as the specific heat can be nicely
fitted by a superposition of double two-level systems, and two fitting
equations for the peak structures are proposed. For the AF-AF-F chain with
strong F interactions, $\chi (T)$ can be well fitted by a polynomial.

For the AF-AF-F chain which is a ferrimagnet, it is observed that the
magnetic curve $m(h)$ reveals some peculiar behaviors where a magnetization
plateau at $m=1/6$ is observed at low temperature $T$, while the plateau is
smeared out when $T$ is increased, and the temperature dependence of the
magnetization $m(T)$ shows various behaviors in different ranges of the
magnetic field. The susceptibility $\chi (T)$ exhibits different behaviors
in different fields: at $h=0$, $\chi (T)$ goes to divergent at lower $T$ and
decreases with increasing $T$ for $J_{AF}/J_{F}\leq 1$; at $h\neq 0$, $\chi
(T)$ shows a low-temperature peak for $J_{AF}/J_{F}=1$ except $h/J_{F}=0.6$
and $1.4$, and decreases at high $T$, where the peak height and position of $%
\chi (T)$ vary under different $h$. The behavior of $\chi T$ against $T$ for
a spin-$5/2$ AF-AF-F chain compound $[Mn_{3}(N_{3})_{6}(bpe)_{3}]$ looks
qualitatively similar to the TMRG result of a spin-$1/2$ AF-AF-F chain.\ The
specific heat $C(T)$ is found to show a double-peak structure, with the peak
height and position varying for different $J_{AF}/J_{F}$ as well as under\
different $h$. The results of $m(T)$, $\chi (T)$ and $C(T)$ show that the
system under interest could be in several different states under different
magnetic fields.

For the F-F-AF chain which is an antiferromagnet, it is found that the
magnetization process shows an interesting behavior, and a magnetization
plateau at $m=1/6$ is also seen at low temperature which is smeared out at
high temperature. $m(T)$ displays different behaviors in different regimes
of the magnetic field, which is further confirmed by the TMRG results of the
susceptibility $\chi (T)$ where a peak is observed at low temperature, but
the peak height and position are found dependent on different ranges of $h$.
The experimental data of $\chi (T)$ for the spin-$1/2$ F-F-AF chain compound 
$3CuCl_{2}\cdot 2dx$ is compared with the corresponding TMRG result, which
is found quite agreement. The specific heat as a function of temperature
shows a double-peak structure, whose shapes change with different $%
J_{AF}/J_{F}$ as well as in different magnetic fields. \ \ \ \ \ \ 

The MSW theory within the Yamamoto scheme as well as the Takahashi scheme is
also applied to study the thermodynamics of the present trimerized quantum
Heisenberg AF-AF-F chain and F-F-AF chain, respectively. Three branch
excitations are found for both systems, say, one gapless excitation spectrum
and two gapful excitation spectra, that come from the competition between
the F and AF interactions. A perfect fitting of $\chi (T)$ and $C(T)$ to a
superposition of double two-level systems could be reasonably understood
within the framework of the MSW theory. It is three branch excitations with
two gaps in the long wave-length limit that are responsible for the peak
structures of the susceptibility and the specific heat observed by the TMRG
calculations. It is found that the MSW theory could to some extent describe
the low-temperature behavior of the systems, and gives only a qualitative
description for $\chi (T)$ and $C(T)$ of AF-AF-F chain and $\chi (T)$ of
F-F-AF chain at high temperature, but has a large deviation for $C(T)$ of
the F-F-AF chain at high temperature.

Finally, we would like to state that the present TMRG and MSW results could
be helpful for understanding the magnetization process as well as the
thermodynamic properties of the AF-AF-F and F-F-AF trimerized quantum spin
chains. As there are few experimental data available for such systems at
present, we expect that more experiments can be performed in near future to
verify our theoretical findings.

\begin{appendix}

\section{MSW THEORY FOR THE TRIMERIZED F-F-AF CHAIN}

The formalism of the MSW theory for the AF-AF-F chain can be found in Ref. 
\cite{Inoue}. In this Appendix, the MSW theory for the F-F-AF chain is
formulated. The Hamiltonian of the 1D spin-$1/2$ F-F-AF chain can be written
as  
\begin{eqnarray}
H &=&\sum\limits_{n=1}^{N}(J_{F}\mathbf{S}_{1n}\cdot \mathbf{S}_{2n}+J_{F}%
\mathbf{S}_{2n}\cdot \mathbf{S}_{3n}+J_{AF}\mathbf{S}_{3n}\cdot \mathbf{S}%
_{4n}  \notag \\
&&J_{F}\mathbf{S}_{4n}\cdot \mathbf{S}_{5n}+J_{F}\mathbf{S}_{5n}\cdot 
\mathbf{S}_{6n}+J_{AF}\mathbf{S}_{6n}\cdot \mathbf{S}_{1n+1}),  \notag
\label{hiffa}
\end{eqnarray}%
where $6N$ is the length of the F-F-AF chain and $J_{F}<0$, $J_{AF}>0$. We
start from the Holstein-Primakoff (HP) transformation 
\begin{eqnarray}
S_{in}^{+} &=&\sqrt{2S-a_{in}^{+}a_{in}}a_{in},  \notag \\
S_{in}^{z} &=&S-a_{in}^{+}a_{in},  \notag \\
S_{jn}^{+} &=&a_{jn}^{+}\sqrt{2S-a_{jn}^{+}a_{jn}},  \notag \\
S_{jn}^{z} &=&-S+a_{jn}^{+}a_{jn},
\end{eqnarray}%
where $S=1/2$, $i=1$, $2$, $3$ and $j=4$, $5$, $6$. The LMSW method treats
the HP transformation up to O($S^{1}$), while the PIMSW method treats is the
HP transformation up to O($S^{0}$) where the interactions between spin waves
are handled perturbatively.

In the conventional spin-wave scheme, the spin deviations in each
sublattice, $\langle a_{ik}^{+}a_{ik}\rangle $,  diverge in the
antiferromagnetic ground state but stay finite in the ferrimagnetic ground
state. In the AF system, the quantum as well as the thermal divergences of
the number of bosons are all suppressed: 
\begin{eqnarray}
0 &=&\sum\limits_{n}\langle
S_{1n}^{z}+S_{2n}^{z}+S_{3n}^{z}-S_{4n}^{z}-S_{5n}^{z}-S_{6n}^{z}\rangle  
\notag \\
&=&6NS-\sum\limits_{k}\sum\limits_{i=1}^{6}\langle a_{ik}^{+}a_{ik}\rangle .
\label{staggt}
\end{eqnarray}

As the F-F-AF is an antiferromagnet, the Takahashi scheme\cite{Taka2} will
be invoked. In order to enforce the constraint (\ref{staggt}), a Lagrange
multiplier $\mu $ is introduced to the Hamiltonian 
\begin{equation*}
\tilde{H}=H+\mu \sum\limits_{k}\sum\limits_{i=1}^{6}a_{ik}^{+}a_{ik}.
\end{equation*}%
The Fourier transformations are defined by 
\begin{eqnarray}
a_{ln} &=&\frac{1}{\sqrt{N}}\sum\limits_{k}e^{ik(n-7/12+l/6)}a_{lk},  \notag
\\
a_{mn} &=&\frac{1}{\sqrt{N}}\sum\limits_{k}e^{-ik(n+7/12-m/6)}a_{mk},
\end{eqnarray}%
where $l=1,2,3$ and $m=4,5,6$. The Hamiltonian, up to the order of $O(S)$,
takes the form of 
\begin{equation*}
H=E_{0}+\sum\limits_{k}\sum\limits_{i=1}^{6}(\omega
_{i}a_{ik}^{+}a_{ik}+\gamma _{i}a_{ik}a_{i+1k}^{+}+\gamma _{i}^{\ast
}a_{ik}^{+}a_{i+1k}),
\end{equation*}%
where $a_{7k}=a_{1k}$, $E_{0}=-2NS^{2}(J_{AF}-2J_{F})$, $\omega _{1,3,4,6}=$$%
S(J_{AF}-J_{F})+\mu $, $\omega _{2,5}=$$-2SJ_{F}+\mu $, $\gamma
_{1,2}=SJ_{F}e^{-i\frac{k}{6}}$, $\gamma _{3}=SJ_{AF}e^{-i\frac{k}{6}}$, $%
\gamma _{4,5}=SJ_{F}e^{i\frac{k}{6}}$, and $\gamma _{6}=SJ_{AF}e^{i\frac{k}{6%
}}$. Via the Bogoliubov transformation 
\begin{eqnarray}
a_{ik} &=&u_{i1}(k)\alpha _{1k}+u_{i2}(k)\alpha _{2k}+u_{i3}(k)\alpha _{3k} 
\notag \\
&&+u_{i4}(k)\alpha _{4k}^{+}+u_{i5}(k)\alpha _{5k}^{+}+u_{i6}(k)\alpha
_{6k}^{+},  \notag \\
a_{jk}^{+} &=&u_{j1}(k)\alpha _{1k}+u_{j2}(k)\alpha _{2k}+u_{j3}(k)\alpha
_{3k}  \notag \\
&&+u_{j4}(k)\alpha _{4k}^{+}+u_{j5}(k)\alpha _{5k}^{+}+u_{j6}(k)\alpha
_{6k}^{+},
\end{eqnarray}%
where $i=1,2,3$ and $j=4,5,6$, the Hamiltonian can be diagonalized as 
\begin{equation*}
H=E_{g}+\sum\limits_{k}\sum\limits_{i=1}^{6}\sigma _{i}E_{ik}\alpha
_{ik}^{+}\alpha _{ik},
\end{equation*}%
where $\sigma _{i}=+$ for $i=1,2,3$ and $-$ for $i=4,5,6$, and $%
E_{g}=E_{0}+\sum\limits_{k}[\sum\limits_{i=1}^{3}\omega
_{i}(\sum\limits_{j=4}^{6}|u_{ij}|^{2})+\sum\limits_{i=4}^{6}\omega
_{i}(\sum\limits_{j=1}^{3}(|u_{ij}|^{2})]$. The coefficients of the
Bogoliubov transformation can be found through equations of motion $i\hbar 
\dot{a}_{ik}=[a_{ik},H]$: 
\begin{equation*}
\left( 
\begin{array}{cccccc}
\omega _{1} & \gamma _{1}^{\ast } & 0 & 0 & 0 & \gamma _{6}^{\ast } \\ 
\gamma _{1} & \omega _{2} & \gamma _{2}^{\ast } & 0 & 0 & 0 \\ 
0 & \gamma _{2} & \omega _{3} & \gamma _{3}^{\ast } & 0 & 0 \\ 
0 & 0 & -\gamma _{3} & -\omega _{4} & -\gamma _{4} & 0 \\ 
0 & 0 & 0 & -\gamma _{4}^{\ast } & -\omega _{5} & -\gamma _{5} \\ 
-\gamma _{6} & 0 & 0 & 0 & -\gamma _{5}^{\ast } & -\omega _{6}%
\end{array}%
\right) \left( 
\begin{array}{c}
u_{1i} \\ 
u_{2i} \\ 
u_{3i} \\ 
u_{4i} \\ 
u_{5i} \\ 
u_{6i}%
\end{array}%
\right) =E_{ik}\left( 
\begin{array}{c}
u_{1i} \\ 
u_{2i} \\ 
u_{3i} \\ 
u_{4i} \\ 
u_{5i} \\ 
u_{6i}%
\end{array}%
\right) .
\end{equation*}%
For a given $k$ and $\mu $, the eigenvalues $E_{ik}$ and eigenvectors $%
(u_{1i},...,u_{6i})$ can be numerically calculated by the driver $ZGEEV.f$
of the $LAPACK$, which is available on the website\cite{LAPACK}.

At finite temperature, the Lagrange multiplier $\mu (T)$ is determined
through the constraint (\ref{staggt}): 
\begin{eqnarray}
6NS &=&\sum\limits_{k}\sum\limits_{j=1}^{6}n_{jk},  \notag \\
n_{lk} &=&\sum\limits_{i=1}^{3}|u_{li}|^{2}\tilde{n}_{ik}+\sum%
\limits_{i=4}^{6}|u_{li}|^{2}(1+\tilde{n}_{ik}),  \notag \\
n_{mk} &=&\sum\limits_{i=1}^{3}|u_{mi}|^{2}(1+\tilde{n}_{ik})+\sum%
\limits_{i=4}^{6}|u_{mi}|^{2}\tilde{n}_{ik},
\end{eqnarray}%
where $l=1,2,3$, $m=4,5,6$, and $n_{jk}=\langle a_{jk}^{+}a_{jk}\rangle _{T}$%
. Defining $\tilde{n}_{ik}\equiv \langle \alpha _{ik}^{+}\alpha _{ik}\rangle
_{T}$, the spin-wave distribution function, we have $\tilde{n}%
_{ik}=[e^{\sigma _{i}E_{ik}/T}-1]^{-1}$ with $\sigma _{i}=+$ for $i=1,2,3$
and $-$ for $i=4,5,6$.

The internal energy and the magnetic susceptibility can be expressed as \cite%
{Taka2} 
\begin{eqnarray}
E &=&E_{g}+\sum\limits_{k}\sum\limits_{i=1}^{6}\sigma _{i}E_{ik}\tilde{n}%
_{ik},  \notag \\
\chi  &=&\frac{1}{3T}\sum\limits_{k}\sum\limits_{i=1}^{6}\tilde{n}_{ik}(%
\tilde{n}_{ik}+1),  \notag
\end{eqnarray}%
where $\sigma _{i}=+$ for $i=1,2,3$ and $-$ for $i=4,5,6$. $E_{ik}$, $\chi $
and $C$ within the LMSW framework can be thus calculated numerically, as
shown in Figs.\ref{msw-ffa}.

For the PIMSW method, the HP transformation is treated, up to the order of O(%
$S^{0}$), such that 
\begin{equation*}
S_{1n}^{+}S_{2n}^{-}\simeq 2Sa_{1n}a_{2n}^{+}-\frac{1}{2}%
a_{1n}^{+}a_{1n}a_{1n}a_{2n}^{+}-\frac{1}{2}a_{1n}a_{2n}^{+}a_{2n}^{+}a_{2n}.
\end{equation*}%
The interactions may be handled in the perturbational way: 
\begin{eqnarray}
a_{1n}^{+}a_{1n}a_{1n}a_{2n}^{+} &\simeq &2\langle a_{1n}^{+}a_{1n}\rangle
_{0}a_{1n}a_{2n}^{+}+2\langle a_{1n}a_{2n}^{+}\rangle _{0}a_{1n}^{+}a_{1n}, 
\notag \\
a_{1n}a_{2n}^{+}a_{2n}^{+}a_{2n} &\simeq &2\langle a_{1n}a_{2n}^{+}\rangle
_{0}a_{2n}^{+}a_{2n}+2\langle a_{2n}^{+}a_{2n}\rangle _{0}a_{1n}a_{2n}^{+}. 
\notag
\end{eqnarray}%
Define $\langle a_{1n}^{+}a_{1n}\rangle _{0}\equiv C_{11}$, $\langle
a_{2n}^{+}a_{2n}\rangle _{0}\equiv C_{22}$ and $\langle
a_{1n}a_{2n}^{+}\rangle _{0}\equiv C_{12}$. We obtain 
\begin{equation*}
\begin{array}{l}
C_{11}=\frac{1}{N}\sum\limits_{k}\langle a_{1k}^{+}a_{1k}\rangle _{0}=\frac{1%
}{N}\sum\limits_{k}(|u_{14}|^{2}+|u_{15}|^{2}+|u_{16}|^{2}),\nonumber \\ 
C_{22}=\frac{1}{N}\sum\limits_{k}\langle a_{2k}^{+}a_{2k}\rangle _{0}=\frac{1%
}{N}\sum\limits_{k}(|u_{24}|^{2}+|u_{25}|^{2}+|u_{26}|^{2}),\nonumber \\ 
C_{12}=\frac{1}{N}\sum\limits_{k}\langle a_{1k}a_{2k}^{+}\rangle _{0}=\frac{1%
}{N}\sum\limits_{k}e^{-i\frac{k}{6}}(u_{11}u_{21}^{\ast }+u_{12}u_{22}^{\ast
}+u_{13}u_{23}^{\ast }),\nonumber%
\end{array}%
\end{equation*}%
and 
\begin{eqnarray}
&&\sum\limits_{n}a_{1n}^{+}a_{1n}a_{1n}a_{2n}^{+}=\sum%
\limits_{k}2(C_{11}a_{1k}a_{2k}^{+}e^{-i\frac{k}{6}}+C_{12}a_{1k}^{+}a_{1k})
\notag \\
&=&\sum\limits_{k}\sum\limits_{i=1}^{6}(2C_{11}e^{-i\frac{k}{6}%
}u_{1i}u_{2i}^{\ast }+2C_{12}|u_{1i}|^{2})\alpha _{ik}^{+}\alpha _{ik}, 
\notag \\
&&\sum\limits_{n}a_{1n}a_{2n}^{+}a_{2n}^{+}a_{2n}=\sum%
\limits_{k}2(C_{12}a_{2k}^{+}a_{2k}+C_{22}a_{1k}a_{2k}^{+}e^{-i\frac{k}{6}})
\notag \\
&=&\sum\limits_{k}\sum\limits_{i=1}^{6}(2C_{22}e^{-i\frac{k}{6}%
}u_{1i}u_{2i}^{\ast }+2C_{12}|u_{2i}|^{2})\alpha _{ik}^{+}\alpha _{ik},
\label{up2s0}
\end{eqnarray}%
where $(...)$ in (\ref{up2s0}) is the up-to-$O(S^{0})$ correction to $E_{ik}$
from the term $\langle S_{1n}^{+}S_{2n}^{-}\rangle $, and the other terms
have similar contributions to $E_{ik}$. The only difference between LMSW and
PIMSW is $E_{ik}$, say, the former is treated up to O($S^{1}$), while the
latter up to O($S^{0}$). Accordingly, $E_{ik}$, $\chi $ and $C$ within the
PIMSW scheme can thus be calculated numerically, as shown in Figs.\ref%
{msw-ffa}.

\end{appendix}

\acknowledgments

The authors are grateful to H. F. Mu, X. Y. Wang, Q. B. Yan and Q. R. Zheng
for helpful discussion and assistance. This work is supported in part by the
National Science Foundation of China (Grant Nos. 90403036, 20490210,
10247002).

\end{document}